# Stochastic multi-step polarization switching in ferroelectrics

Y.A. Genenko[1*], R. Khachaturyan[1], J. Schultheiß[2], A. Ossipov[3], J. E. Daniels[4], and J. Koruza[2]

[1]Institute of Materials Science, Technische Universität Darmstadt, Otto-Berndt-Str. 3, 64287 Darmstadt, Germany

[2]Institute of Materials Science, Technische Universität Darmstadt, Alarich-Weiss-Straße 2, 64287 Darmstadt, Germany

[3]School of Mathematical Sciences, University of Nottingham, University Park, Nottingham NG7 2RD, UK

[4]School of Materials Science and Engineering, UNSW Sydney, NSW, 2052, Australia

## Abstract

Consecutive stochastic 90° polarization switching events, clearly resolved in recent experiments, are described by a new nucleation and growth multi-step model. It extends the classical Kolmogorov-Avrami-Ishibashi approach and includes possible consecutive 90°- and parallel 180°-switching events. The model predicts the results of simultaneous time-resolved macroscopic measurements of polarization and strain, performed on a tetragonal Pb(Zr,Ti)O$_3$ ceramic in a wide range of electric fields over a time domain of five orders of the magnitude. It allows the determination of the fractions of individual switching processes, their characteristic switching times, activation fields, and respective Avrami indices.

*Corresponding author: genenko@mm.tu-darmstadt.de

1. Introduction

Polarization switching driven by an applied electric field is a fundamental process in ferroelectrics involving thermally activated nucleation and growth of reversed polarization domains. Understanding the kinetics of this process is important for many applications, particularly, for ferroelectric memories (FERAM) [1]. Previously, macroscopic polarization switching kinetics was described by stochastic models, such as the classical Kolmogorov-Avrami-Ishibashi (KAI) model based on the concept developed to describe melt solidification [2] and assuming random and statistically-independent nucleation and growth of reversed polarized domains in a uniform medium [3,4]. This stochastic model works well for some single crystals [5-7] but performs unsatisfactorily when applied to polycrystalline ferroelectric films [8-10] or bulk polycrystalline ceramics [11,12]. A range of intrinsic physical features of ferroelectrics are missing in the KAI approach.

The KAI model assumes only a single characteristic switching time for the whole macroscopic system. Introducing a statistical distribution of switching times, characterizing different regions of such a nonuniform system as a polycrystalline solid, helped to improve the characterization of switching kinetics in ferroelectric films [8,13-14] and bulk ceramics [15,16]. The regions are distinguished based on different local electric field amplitudes, originating from the random crystallographic orientations of the grains [16]. Though this model extension provided rather accurate description of polarization response in a range of ferroelectric materials [7,13,14,16-20] another important feature still remained missing, namely the feedback due to depolarization fields emerging during the polarization reversal of individual regions [21-26]. In model simulations [27,28] interaction of different switching regions via the depolarization fields was shown to play an important role in uniform systems providing highly coherent switching in single crystals at long spatial ranges. In contrast, in polycrystalline media the emerging depolarization fields appear to be effectively screened by adapting local bound charges, as disclosed by recent simulations using the self-consistent mesoscopic switching model [29]. Thus, the switching of different regions in a ceramic can still be considered as statistically independent with regard to electrical interactions. Correlations in polarization switching, observed between tens of grains in bulk samples [30,31] or up to thousand grains in films [32-34], seem to be related to the elastic rather than to the electric interactions. Therefore, the original definition of different regions has to be extended to take elastic interactions into account. Beyond the deficiencies listed above, a common shortcoming of all mentioned statistical concepts of polarization switching [2-4,8,13-17,24,26] is that individual random switching events are assumed to occur statistically independent and parallel to each other. In reality, however, some events occur in succession as, for example, consecutive 90°- or, generally, non-180°-switching events. This consideration is crucial for multiaxial ferroelectrics, which are the most widely-used group of ferroelectric materials. Such two-step polarization reversals were observed by *in situ* x-ray diffraction measurements [35] and ultrasonic investigations [36]. Respective characteristic times for two distinct and sequential domain reorientation steps were determined [37]. Furthermore, some reports suggest that contributions from 180°-switching events during the reversal process cannot be excluded [38,39]. In order to distinguish between both contributions, the macroscopic strain of the polycrystalline sample should be measured simultaneously with the switched polarization.

The present work is devoted to the development of a model to describe polarization and strain switching dynamics with consecutive stochastic switching events and its implementation for fitting of macroscopic measurements of a bulk ferroelectric ceramic. In Section 2, the classical KAI model is extended by including two sequential non-180° polarization reorientation steps and a parallel 180° switching event which we call a multi-step stochastic mechanism (MSM) model. Additionally, a relation between the time-dependent strain and polarization is derived. In Section 3, polarization and strain switching experiments over a time domain from $10^{-4}$ to $10^{1}$ s are presented for a range of applied electric field values. The experimental results are analyzed and discussed in Section 4 based on the concepts from Section 2. Finally, the results are concluded in Section 5.

## 2. Theory of consecutive stochastic polarization switching processes

### A. Extension of the KAI model to consecutive switching events

Let us first consider a consecutive 90°-switching process in a polycrystalline ferroelectric in the spirit of the KAI model [2-4]. It is assumed, for simplicity, that polarization may adopt only directions parallel or perpendicular to the electric field, which is applied along the $z$ axis of the Cartesian coordinate system $(x,y,z)$, see Fig. 1.

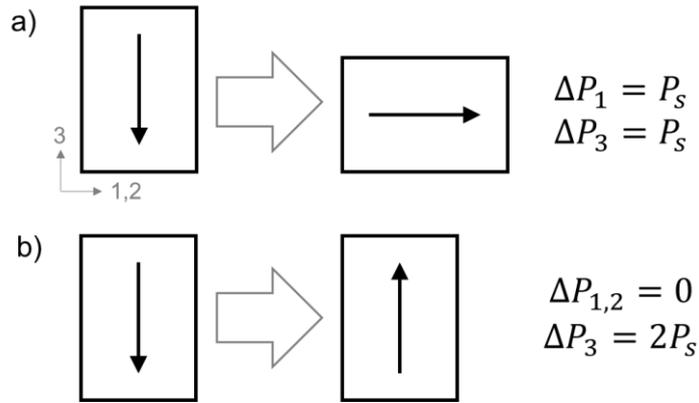

Fig. 1: Changes in polarization due to an idealized a) 90° and a b) 180° switching event.

In the initial state, the system is assumed to be uniformly polarized downwards, exhibiting a saturation polarization $-P_s$. When a reversed (positive) field is applied, the local polarization may experience two sequential 90°-switching events with respective nucleation rates per unit time and unit volume $R_1$ and $R_2$. We first consider the nucleation of switched domains according to the first process after the application of the electric field upwards at time $t > 0$. When an unconstrained domain emerges at some point $B$ at a time $\tau > 0$ it is supposed to grow with a constant (field dependent) velocity $v_1$ so that its "spherical" volume reaches the value

$$\Omega_1(t,\tau) = C_1 \left[ v_1(t-\tau) \right]^{n_1} \qquad (1)$$

by the time $t > \tau$, where $n_1$ is the spatial dimensionality of the domain and $C_1$ is an appropriate numerical coefficient. Here, a possible initial nucleus size is neglected. Let us evaluate the probability $q_1(t)$ for a point $A$ not to be comprised by a switched area of some domain. To this end, following Ishibashi and Takagi [4], let us construct a "spherical" volume $\Omega_1(t,\tau)$ around the point $A$ (see Fig. 2). If the nucleation point $B$ were present in the latter volume, the switched domain would cover the point $A$ by the time $t$. Thus, the probability that no nucleus emerges in the volume $\Omega_1$ around $A$ during the time interval $(\tau, \tau + \Delta\tau)$ equals $1 - R_1(\tau)\Omega_1(t,\tau)\Delta\tau$. The time is now discretized in short steps $\Delta\tau, 2\Delta\tau, ...i\Delta\tau, ...$ from zero until the time $t = N\Delta\tau$. The time of the domain appearance is indicated as $\tau = i\Delta\tau$ with the index *i* varying from zero to *N*. Then the probability that the point *A* is not covered by the switched area by the time *t* results as the product of such probabilities in all elapsed intervals:

$$q_1(t) = \prod_{i=0}^{N}\left[1 - R_1(i\Delta\tau)\Omega_1(N\Delta\tau, i\Delta\tau)\Delta\tau\right]. \qquad (2)$$

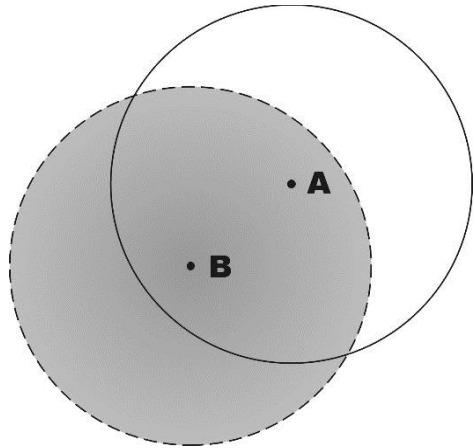

Fig. 2. Scheme of probability calculation in the KAI model. Dark area presents a growing reversed domain nucleated at point B and covering point A.

The logarithm of Eq. (2) brings about the sum which transforms to the integral when $\Delta\tau \to 0$:

$$\ln q_1(t) = -\sum_{i=0}^{N} R_1(i\Delta\tau)\Omega_1(N\Delta\tau, i\Delta\tau)\Delta\tau \Big|_{\Delta\tau\to 0} \to -\int_0^t d\tau\, R_1(\tau)\Omega_1(t,\tau) \qquad (3)$$

so that

$$q_1(t) = \exp\left[-\int_0^t R_1(\tau)\Omega_1(t,\tau)d\tau\right]. \qquad (4)$$

Now we consider a sequence of two switching events according to the first and then to the second 90°-switching event. First let us evaluate a probability $p_1(t_1, \Delta t_1)$ of switching according to the first mechanism (first 90° switching event) within the interval $(t_1, t_1 + \Delta t_1)$, which can be derived from the relation

$$q_1(t_1 + \Delta t_1) = q_1(t_1)\left[1 - p_1(t_1, \Delta t_1)\right]. \qquad (5)$$

By expansion of the left hand side up to the first order in $\Delta t_1$ and assuming the time independent $R_1$ one finds, using Eq. (4),

$$\ln q_1(t_1 + \Delta t_1) = \ln q_1(t_1) - R_1\Omega_1(t_1, 0)\Delta t_1. \qquad (6)$$

From comparison with Eq. (5) it is apparent that

$$p_1(t_1, \Delta t_1) = R_1\Omega_1(t_1, 0)\Delta t_1. \qquad (7)$$

For the probability not to switch according to the second mechanism (second 90° switching event) we obtain similar to (4)

$$q_2(t) = \exp\left[-\int_0^t R_2(\tau)\Omega_2(t,\tau)d\tau\right] \tag{8}$$

with $\Omega_2(t,\tau) = A_2\left[v_2(t-\tau)\right]^{n_2}$ \hfill (9)

where parameters $A_2, v_2$, and $n_2$ characterize the second switching process in analogy to the first one. The probability to switch once according to the first mechanism within the interval $(t_1, t_1 + \Delta t_1)$ and not to switch anymore until the time $t$ is then

$$q_1(t_1)R_1\Omega_1(t_1,0)\Delta t_1 q_2(t-t_1). \tag{10}$$

Finally, the total probability to switch once according to the first mechanism and not to switch anymore until time $t$ is obtained by summation over all possible intervals $(t_1, t_1 + \Delta t_1)$ as

$$L_1(t) = \int_0^t dt_1 q_1(t_1)R_1\Omega_1(t_1,0)q_2(t-t_1). \tag{11}$$

The total probability to switch firstly according to the first mechanism and secondly according to the second mechanism until time $t$ reads apparently as

$$L_2(t) = \int_0^t dt_1 q_1(t_1)R_1\Omega_1(t_1,0)\left[1 - q_2(t-t_1)\right]. \tag{12}$$

By substituting Eqs. (1,4,8) into Eqs. (11,12) one finds general forms

$$L_1(t) = \frac{\alpha}{\tau_1}\int_0^t dt_1 \left(\frac{t_1}{\tau_1}\right)^{\alpha-1}\exp\left[-\left(\frac{t_1}{\tau_1}\right)^\alpha - \left(\frac{t-t_1}{\tau_2}\right)^\beta\right]$$

$$L_2(t) = \frac{\alpha}{\tau_1}\int_0^t dt_1 \left(\frac{t_1}{\tau_1}\right)^{\alpha-1}\exp\left[-\left(\frac{t_1}{\tau_1}\right)^\alpha\right]\left\{1 - \exp\left[-\left(\frac{t-t_1}{\tau_2}\right)^\beta\right]\right\} \tag{13}$$

where the switching times $\tau_1$ and $\tau_2$ for the first and the second processes are defined by the geometrical and kinetic characteristics of the growing domains, which can in principle be different. These parameters, as well as the exponents $\alpha$ and $\beta$, will be used to fit experimental data on the time dependent polarization and strain.

Let us define a mean polarization variation due to all 90°-reorientation events by $P_{90}$. Then the total polarization change all over the system due to the two sequential polarization variations by $P_{90}$ amounts to

$$\Delta p(t) = P_{90} L_1(t) + 2 P_{90} L_2(t) \tag{14}$$

or, equivalently,

$$\Delta p(t) = 2 P_{90} \left\{ 1 - \exp\left[ -\left(\frac{t}{\tau_1}\right)^\alpha \right] \right\} - P_{90} L_1(t) \tag{15}$$

noting that

$$L_2(t) = 1 - \exp\left[ -\left(\frac{t}{\tau_1}\right)^\alpha \right] - L_1(t). \tag{16}$$

Unfortunately, the integrals in Eqs. (13) cannot be generally solved in a closed form for arbitrary $\alpha$ and $\beta$. Their qualitative behavior can, however, be comprehended from a simple particular case $\alpha = \beta = 1$. For this choice

$$L_1(t) = \frac{\tau_2}{\tau_2 - \tau_1} \left( e^{-t/\tau_2} - e^{-t/\tau_1} \right) \quad \text{and} \quad L_2(t) = 1 + \frac{\tau_1}{\tau_2 - \tau_1} e^{-t/\tau_1} - \frac{\tau_2}{\tau_2 - \tau_1} e^{-t/\tau_2}. \tag{17}$$

Considering a typical situation with $\tau_1 \ll \tau_2$ [37], $L_1(t) \cong 1 - e^{-t/\tau_1}$ for $0 < t < \tau_1$ and $L_1(t) \cong e^{-t/\tau_2}$ for $\tau_1 \ll t < \tau_2$. Similarly, for arbitrary indices $\alpha$ and $\beta$ the function $L_1(t)$ first increases on the time scale of $\tau_1$ and then decreases on the time scale of $\tau_2$, vanishing asymptotically.

### B. Combination of consecutive 90°- and parallel 180°-switching events

Analysis of the experimental data by many authors shows that the application of an external field can drive the motion of both 180° and non-180° domain walls [35,36,38,39]. As will be shown later, also in our case polarization and strain measurements can only be comprehended when introducing additionally simultaneous 180°-switching events (Fig. 1b). Thereby, the total switched polarization will be denoted $2P_s$ and can be directly determined from experiment. The contributions of both consecutive 90°-switching events are assumed equal to $P_{90} = P_s \eta$ with a positive $\eta < 1$ presenting the fraction of 90°-events and used further as a fitting parameter. The mean amplitude of the parallel 180°-switching events is then given by $P_{180} = P_s(1-\eta)$ so that $P_{180} + P_{90} = P_s$. The temporal polarization variation is then presented by extension of Eq. (15) as

$$\Delta p(t) = 2P_s \eta \left\{ 1 - \exp\left[-\left(\frac{t}{\tau_1}\right)^\alpha\right]\right\} - P_s \eta L_1(t) + 2P_s(1-\eta)\left\{1 - \exp\left[-\left(\frac{t}{\tau_3}\right)^\gamma\right]\right\} \quad (18)$$

where the first two terms represent the contributions from 90°- and the third term contributions from 180°-switching events with the corresponding switching time $\tau_3$ and the Avrami exponent $\gamma$. In addition to the aforementioned parameters, the latter two will also be used for fitting the time-dependent polarization reversal and strain data.

## C. Relation between simultaneous polarization and strain in a system with stochastic consecutive 90°- and parallel 180°-events

To prove the consistency of the theory advanced above, we derive here the contribution to the strain tensor $S_{ij}$ directly from the variation of polarization $p_n$. To this end one can use a relation between the strain and the polarization derived from electrostriction, valid for any solid [40,41],

$$S_{ij} = Q_{ijmn} p_m p_n \qquad (19)$$

with the electrostriction tensor $Q_{ijmn}$, if no stress is applied to the system. For ferroelectrics with a cubic parent phase, the piezoelectric contribution results from Eq. (19) when the spontaneous polarization **P** is singled out as

$$p_n \cong P_n + \varepsilon_0 \varepsilon_{nm} E_m \qquad (20)$$

with the permittivity of vacuum $\varepsilon_0$ and the relative permittivity of the ferroelectric $\varepsilon_{nm}$. By substitution of Eq. (20) into Eq. (19) and neglecting a small quadratic field contribution [42] one obtains

$$S_{ij} \cong Q_{ijmn} P_m P_n + d_{ijk} E_k \qquad (21)$$

whereby the general equation for the piezoelectric coefficient is used [41]

$$d_{ijk} = 2\varepsilon_0 \varepsilon_{km} Q_{ijml} P_l \,. \qquad (22)$$

Using the Voigt notations [43] these formulas can be specified for the direction $z$ of the macroscopic strain measurements as

$$S_3 = Q_{11} P_3^2 + Q_{12}\left(P_1^2 + P_2^2\right) + 2\varepsilon_0 \varepsilon_{33} Q_{11} E_3 P_3 \qquad (23)$$

Note, that the $P_3$ component is changed by 180°, as well as 90° events and thus the piezoelectric part contains contributions from the intrinsic lattice expansion and domain switching processes.

The quantity measured in the experiment is the field-driven variation of the strain $\Delta S_3 = S_3 - S_3^0$, whereby $S_3^0$ is the remanent strain of a sample fully polarized downwards. In the considered model, polarization components can only adopt values $P_n = \pm P_s$ or 0 at any time so that $P_1^2 + P_2^2 + P_3^2 = P_s^2$. Thus, in the initial state we assume $P_3 = -P_s$ everywhere and $S_3^0 = Q_{11} P_s^2$. Then the strain variation can be expressed as

$$\Delta S_3 = (Q_{12} - Q_{11})(P_1^2 + P_2^2) + 2\varepsilon_0 \varepsilon_{33} Q_{11} E_3 P_3. \tag{24}$$

The 180°-switching processes fully contribute to the variation of polarization $P_3$ (see figure 1a) and thus only change the strain by the linear term in Eq. (24). In contrast, the first and the second 90° switching events rotate the unit cell by 90°, thus contributing to the variation of the strain by both terms in Eq. (24). The squared transverse polarization $P_1^2 + P_2^2 = P_\perp^2$ adopts by the first 90°-switching event a value $P_s^2$ resulting in the maximum possible spontaneous strain $\Delta S_{\max} = (Q_{12} - Q_{11}) P_s^2$. It is also important to note that the first 90°-rotation of polarization contributes to the strain by $\Delta S_{\max}$ and the second 90°-rotation changes it by $-\Delta S_{\max}$, so that two consequent 90° rotations are equivalent to one 180° switching and thus in sum cause no variation of the strain by the quadratic term, but by the linear term in Eq. (24). Using the switching probabilities derived in the previous section 2B, the averaged strain variation can now be expressed as

$$\Delta S_3(t) = \Delta S_{\max} \eta L_1(t) + 2\varepsilon_0 \varepsilon_{33} Q_{11} E_3 (\Delta p(t) - P_s) \tag{25}$$

with functions $L_1(t)$ and $\Delta p(t)$ given by Eqs. (13) and (18), respectively. Note that both formulas for polarization, (18), and strain, (25), present averaging over the whole system and neglect electric and elastic interactions [44] between different switching regions during the polarization reversal.

## 3. Experimental work

Bulk, polycrystalline Pb$_{0.985}$La$_{0.01}$(Zr$_{0.475}$Ti$_{0.525}$)O$_3$ ceramics were prepared by a mixed-oxide route [45]. The switched polarization and the macroscopic strain were measured simultaneously. The samples were poled in direction downward with an electric field of 3 kV/mm for 20 s. After a wait time of 100 s, a 10 s pulse switching field $E_{Sw}$ was applied opposite to the poling direction. In order to realize a sharp high voltage (HV) pulse rise of 115 ns (rise time up to 75% of the maximal voltage), a buffer capacitor, which was charged by a high voltage source (Trek Model 20/20C, Lockport, NY, USA), was combined with a commercial fast HV transistor switch (HTS 41-06-GSM, Behlke GmbH, Kronberg, Germany) [46]. The charge was monitored by measuring the voltage drop across a reference capacitor (WIMA MKS4, Wima, Mannheim, Germany), while the macroscopic displacement of the sample was simultaneously measured by an optical displacement senor (D63, Philtec Inc., Annapolis, MD, USA) with a time resolution of $10^{-4}$ s.

Fig. 3 displays the time-dependent data of polarization and strain for various switching fields $E_{SW}$. Note that the leakage current and the dielectric displacement were subtracted in the presented polarization of $\Delta p$ data. The variation of the strain, $\Delta S_3$, starts in all measurements from zero which is not explicitly seen in the plot since the data below $10^{-4}$ s are not available.

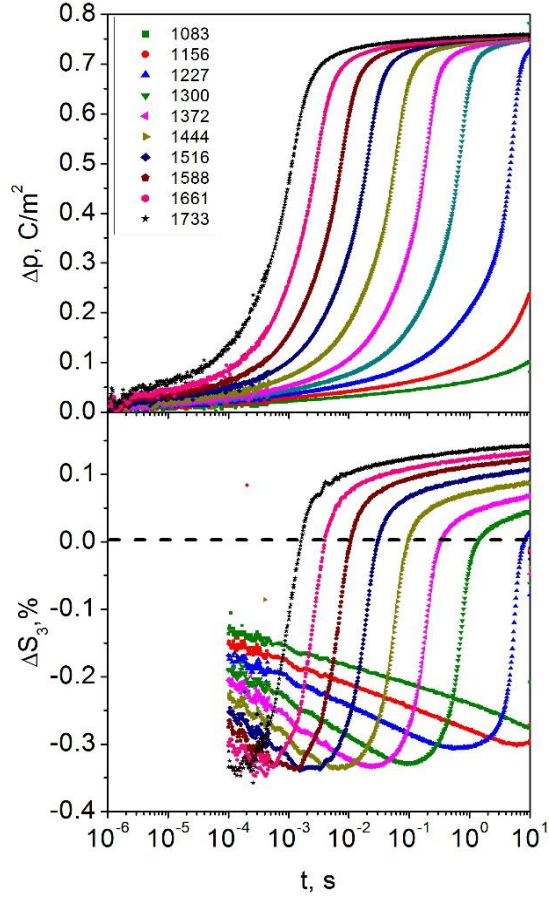

Fig. 3. (a): Results of the simultaneous dynamic measurements of a) switched polarization, $\Delta p$, and (b) strain, $\Delta S_3$, of a polycrystalline PZT ceramic. The curves were measured at different applied fields, $E_{SW}$, as indicated by the inset values in kV/mm. The dashed line represents the initial value $\Delta S_3 = 0$ for $E_{SW} = 0$, which is related to the remanent strain, $S_3^0$.

Electrostrictive coefficient, $Q_{11}$, and large signal permittivity, $\varepsilon_{ls}$, were determined by fitting the high-field part of bipolar polarization and strain measurements. For the former, equation (18) was fitted as $S_3 = Q_{11} P_3^2$ [47] to the data in Fig. 4(a) and $Q_{11}$ was determined as $Q_{11} = 0.046 \, \text{m}^4 / \text{C}^2$, which is close to the value of $0.044 \, m^4/C^2$ reported for PZT ceramic at the tetragonal side of the morphotropic phase boundary [41,47]. The large field permittivity, $\varepsilon_{ls}$, was calculated by normalization of the derivative of the polarization with respect to the electric field: $\varepsilon_{ls} = (dP/dE)/\varepsilon_0$ [48]. As shown in Fig. 4(b) a saturated value of about $\varepsilon_{ls} = 3 \cdot 10^3$ was obtained.

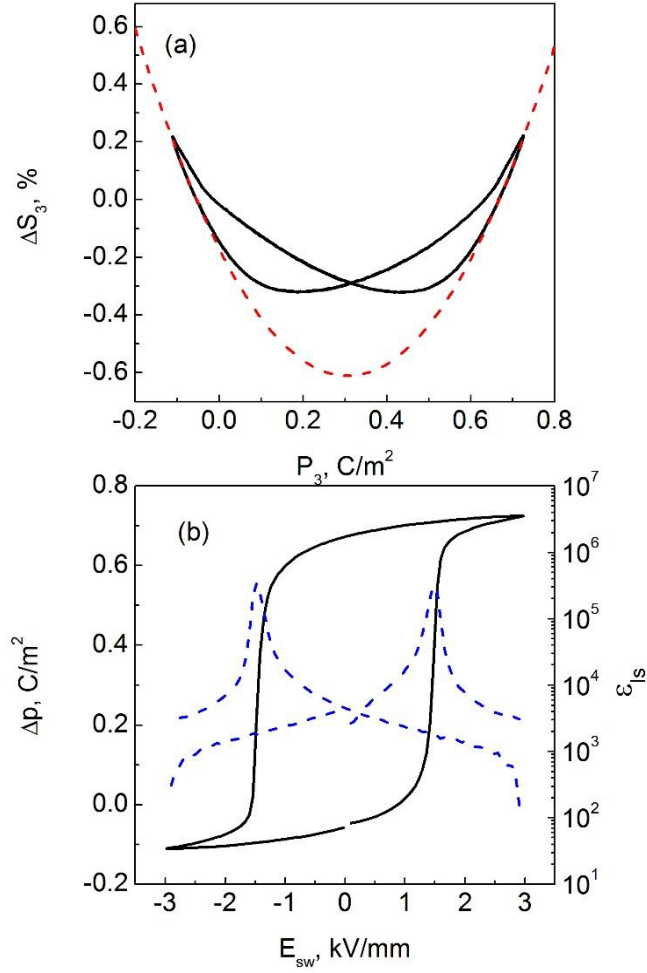

Fig. 4. Experimental determination of the (a) electrostrictive coefficient, $Q_{11}$, and (b) large signal permittivity, $\varepsilon_{ls}$, from bipolar polarization and strain loops.

## 4. Analysis and discussion of experimental results

Since Eq. (18) contains a smaller number of fitting parameters than Eq. (25), the polarization-time curves (Fig. 3(a)) were fitted first. This fitting defines in the first iteration the parameters η, $\tau_1, \tau_2, \tau_3$, α, β, and γ. These parameters were then kept constant while fitting the corresponding strain-time curves using $\Delta S_{max}$ and $\varepsilon_{33}$ in Eq. (25) as variable parameters. Though this procedure provided at once a satisfactory agreement with both polarization-time and strain-time experimental curves, a few further iteration steps were carried out because the description of the strain-time curve turned out to be more sensitive for the parameters $\tau_1, \tau_2$.

Having the set of parameters established enables us to describe the time-dependency of the macroscopic strain and polarization, as shown in Fig. 5 for an applied field strength of 1.588 kV/mm.

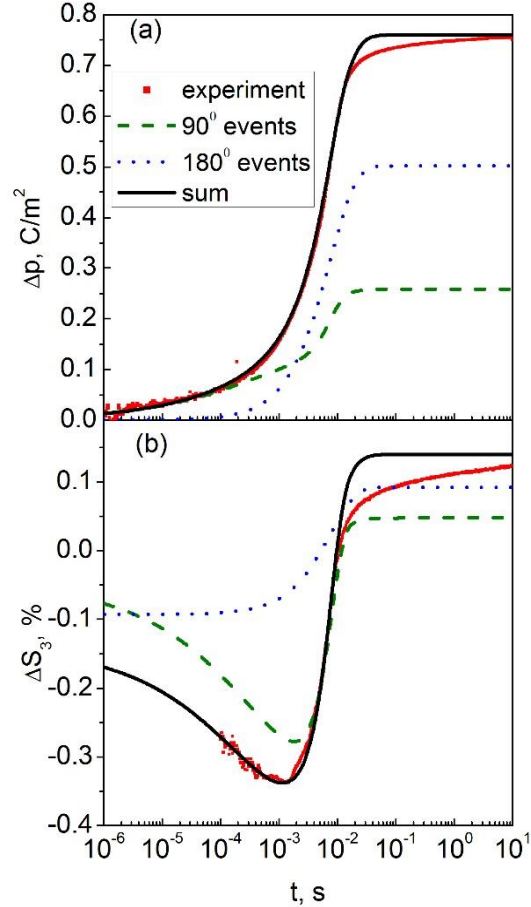

Fig. 5. Polarization (a) and strain (b) variation with time at the applied field of 1.588 kV/mm. Experimental curves are shown by symbols, separated theoretical contributions from 90° switching events by dashed lines, contributions from 180° switching events by dotted lines, and their sum by solid lines.

The above described fitting procedure was performed separately for all data sets obtained experimentally for each value of the electric field, $E_{Sw}$, shown in Fig. 3. The materials characteristics $P_s, Q_{11}, \Delta S_{max}, \varepsilon_{33}$ were thereby kept constant all over the studied field region. The shares of 90° and 180° switching processes were found to be field independent with $\eta = 0.34$ within the considered field range 1.1-1.7 kV/mm. However, even though similar 180° and non-180° switching shares were previously reported [38,39], it should be noted that 180° switching can also happen strain-free by two statistically dependent, coherent 90° events (Fig. 6), as suggested by Arlt [49].

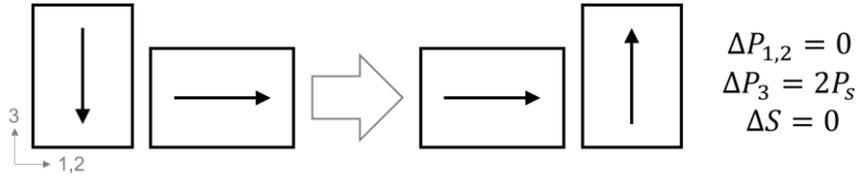

Fig. 6. Example of two coherent 90° switching events which do not change the overall strain but contribute to the $P_3$ polarization component, similar to a 180° switching event.

The results of fitting are exemplarily presented in Fig. 7 for representative field values. As is seen, Equations (18) and (25) well approximate both polarization-time and strain-time curves in the short and intermediate time regions, whereby the latter is identified by the maximum switching rate. Furthermore, these formulas allow the prediction of polarization and strain dependences beyond the observed time interval if appropriate fitting was performed at shorter times, as is demonstrated in Fig. 7 (a) and 7 (b).

Nevertheless, theoretical curves (solid lines) notably deviate from experimental ones (symbols) at the later stages of switching, when approaching the saturated polarization. As compared to the experiment, the analytical calculations exhibit a sharper step-like behavior, typical for the classical KAI-concept [5,7,9] and observed in single crystals [6,7]. In ceramics, however, a more dispersive behavior is typical at longer switching times [7-20]. Physical reasons of this behavior are still disputed and may be attributed to a creep-like domain wall movement of ferroelastic domains [50] related to a broad distribution of the switching times [8]. This could be explained by the inhomogeneous field mechanism (IFM) model [15,16], which derives a wide statistical distribution of switching times from the non-uniformly distributed electrical field in random systems, such as polycrystalline ferroelectrics. Over and above, the statistical field distribution does not remain fixed in the course of polarization reversal and develops due to varying depolarization fields [21-26]. As was shown in a recent work [29], the field distribution is widening with the polarization increase and mostly affects the poling process at later stages approaching the saturation. This could explain the discrepancy between experimental and fitting curves. In the current model, however, we would like to focus on the statistical explanation of sequential 90°-switching processes. Introduction of distributed switching times could improve fitting of the experiment data, but this would make a model more sophisticated, less transparent, and would exceed the scope of this work.

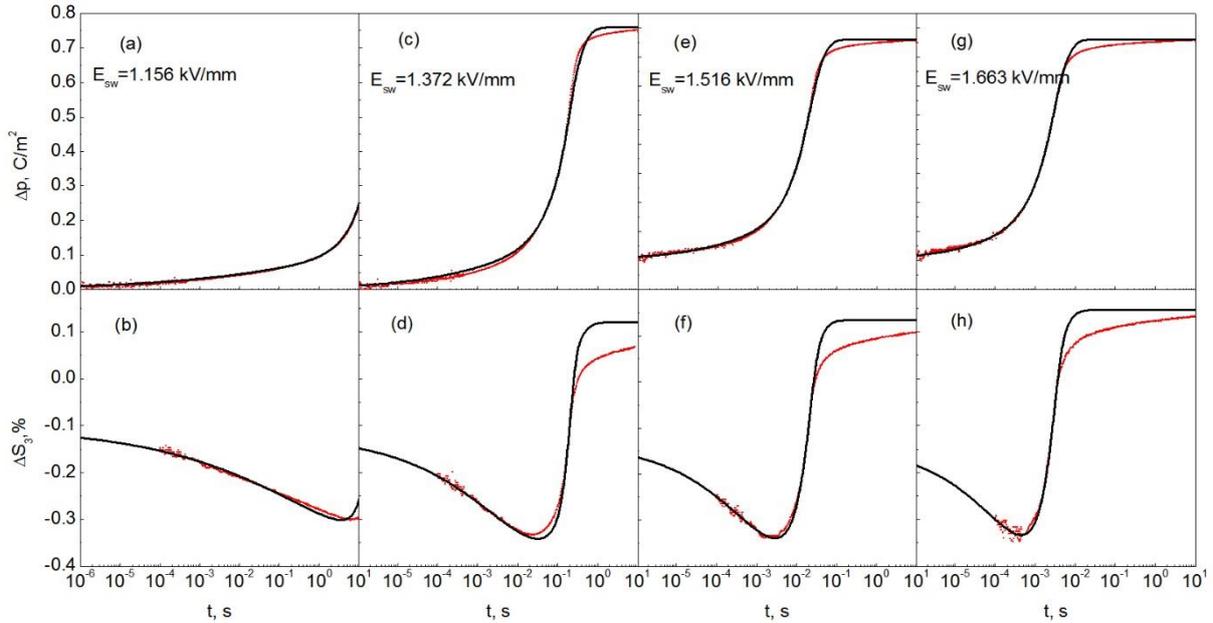

Fig. 7. Variation of the polarization (a,c,e,g) and strain (b,d,f,h) with time at different field values in kV/mm as indicated in the plots. Experimental curves are shown by symbols and fitting curves by solid lines.

The neglected field and consequently time distributions are presumably also responsible for non-integer values of the Avrami exponents (see Fig. 8), which are well-known from publications trying to explain a dispersive polarization response within the KAI approach [11,51-54]. However, the variation of the Avrami exponents might also have a physical meaning. Thus, an abrupt variation of the dimensionality of growing reversed domains from 3D towards 2D towards 1D domain geometry was identified in polarization kinetics experiments on PZT thin films and simulations [55]. This can be related to the jump in the Avrami index $\beta$ due to coalescence of numerous small domains to a large stripe-like ones at higher electric fields.

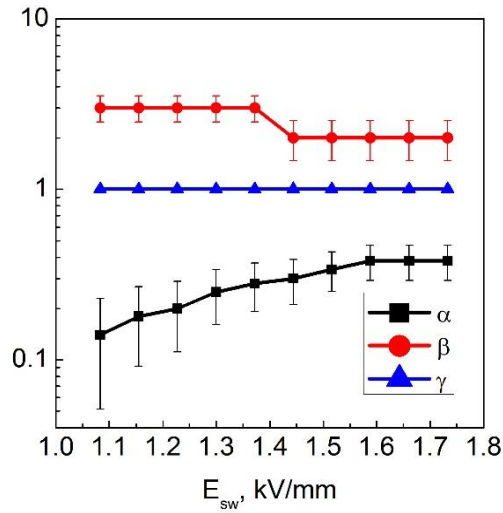

Fig. 8. The Avrami exponents obtained by fitting at different field values. Symbols indicate the best fit values and error bars their standard deviations.

The values of characteristic switching times, extracted from the dynamic curves of Fig. 7, are shown in Fig. 9. The switching times $\tau_2$ and $\tau_3$ exhibit the Merz law behavior [56], $\tau = \tau_0 \exp(E_a / E_{SW})$. Activation field, $E_a$, values for these two events were calculated to be about 33 kV/mm. The field dependence of $\tau_1$ could not be described by the Merz law with a single activation field value over the entire field range, as previously reported for some ceramics and temperature regimes [17,19]. However, it is obvious that the activation energy for the first 90°-switching event is the smallest among the three events and is near 27.7 kV/mm in the high-field region. The lower value of the activation field for the first switching may be assigned to the promoting effect of the residual stresses, suggested by x-ray diffraction studies [30,37]. The fitting has revealed a relative permittivity value of $2.85 \cdot 10^3$ being field independent as expected for this field range [57]. This value is comparable with the experimentally measured value of $3 \cdot 10^3$ evaluated as indicated above in Section 3.

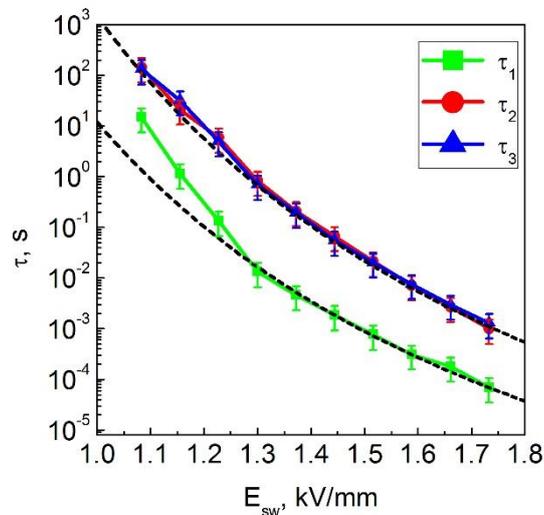

Fig. 9. Characteristic switching times, extracted by fitting and approximated by the Merz law, with $\tau_0 = 0.8 \cdot 10^{-11}$ s, and activation fields 33 kV/mm and 27.7 kV/mm for the upper and lower curves, respectively.

The maximum strain $\Delta S_{max}$ was found to be about -1%, assuming that all switching occurs by 90°, as shown in the idealized model in Fig. 1(b). This parameter can, in principle, be estimated independently using data from mechanical loading (ferroelastic) experiments, whereby the poled sample is uniaxially compressed in the *z*-direction. The value reported for the maximum strain in such conditions was around -0.68% [58]. Real materials deviate from the idealized model shown in Fig. 1(b), because the possible polarizations of grains are specified by their crystallographic orientations, which are randomly distributed. Additionally, in bulk polycrystals, domains can interact across grain boundaries leading to longer length-scale coupling of domain dynamics [30,31,59]. The c-axis directional distribution should be taken into account and the simplified formula for $\Delta S_{max}$ has to be generalized to

$$\Delta S_{\text{max\_real}} = (Q_{12} - Q_{11})P_s^2 <(\sin^2\theta)>_{max} \qquad (26)$$

where $<\sin^2\theta>_{max}$ defines a maximum possible value for $(P_x^2 + P_y^2)/P_s^2$ compatible with the tetragonal symmetry of grains when polarization tries to avoid *z*-direction. Thus the parameter of the simplified model is related to the observed value by $\Delta S_{max} = \Delta S_{\text{max\_real}}/<(\sin^2\theta)>_{max}$. It is known that for tetragonal symmetry $<(\cos^2\theta)>_{max}=0.701$, which defines a lower limit for $<\sin^2\theta>$ as $1-<(\cos^2\theta)>_{max}= 0.299$. The maximum limit $<\sin^2\theta>_{max}$ is expected to be close to $<(\cos^2\theta)>_{max}$. The fitting value $\Delta S_{max}$ of -1% corresponds to $<(\sin^2\theta)>_{max} \approx 0.6$. From the maximum strain the other electrostriction coefficient can be evaluated as $Q_{12} = Q_{11} - \Delta S_{max}/P_S^2 = 0.021 \ (\text{m}^4/\text{C}^2)$.

## 5. Conclusions

We have developed a multi-step stochastic mechanism (MSM) model of the field-driven polarization reversal in ferroelectric ceramics. Similar to the classical KAI approach, this model assumes statistically independent, non-correlated polarization switching region by region, neglecting both elastic and electric interaction between the switching regions. However, in contrast to the classical KAI consideration, the model includes two parallel channels of switching: a 180°-polarization reversal and a sequential two-step 90°-switching events. Application of the model to the experimental results of simultaneous macroscopic measurements of polarization and strain, over a wide time window performed at different applied fields, allowed determination of such characteristics of the switching processes as their field-dependent characteristic times and Avrami indices. Other parameters extracted from fitting of the experimental data, such as the maximum spontaneous strain and dielectric permittivity, are in agreement with independently measured values. A very important result of the analysis is the share of 90°-switching events, which appears to be field independent in the used field region 1.1-1.7 kV/mm and equals $\eta$=0.34. This value, however, should be treated with care because it counts only the statistically independent 90°-switching events included in our stochastic model. For example, such statistically dependent, coherent 90°-switching events that do not contribute to the strain (see Fig. 9), cannot be identified in this approach and thus they would appear as parallel 180°-switching events.

**Acknowledgements**

This work was supported by the Deutsche Forschungsgemeinschaft (DFG) Grants Nos. GE 1171/7-1 and KO 5100/1-1. JED acknowledges financial support from an Australian Research Council Discovery Projects DP120103968 and DP130100415. M. Weber is acknowledged for the construction of the electrical HV setup. Dr. H. Kungl is acknowledged for the preparation of the sample.